\documentclass[aps,preprint,amsmath,amssymb,amsfonts,nofootinbib]{revtex4}
\usepackage{epsfig}
\usepackage{graphicx}
\usepackage{dcolumn}
\usepackage{bm}
\usepackage{amsthm}
\usepackage{amsmath}
\usepackage{color}

\newcommand{\del}{\partial}

\newcommand{\beq}{\begin{equation}}
\newcommand{\eeq}{\end{equation}}
\newcommand{\bea}{\begin{eqnarray}}
\newcommand{\eea}{\end{eqnarray}}

\newcommand{\half}{\frac{1}{2}}
\newcommand{\p}{\partial}
\newcommand{\ep}{\epsilon}
\newcommand{\bep}{\bar{\epsilon}}
\def\a{\dot{\alpha}}
\def\ee{\varepsilon}
\newcommand{\dslash}{\slash{\!\!\!\del}}
\newcommand{\sla}[1]{\slash\!\!\!#1}

\begin{document}
\begin{flushright}
\end{flushright}
\bigskip
\bigskip

\title{The Super-Higgs Mechanism in Fluids}
\author{Karim Benakli$^{1,2}$}
\author{Yaron Oz$^3$}
\author{Giuseppe Policastro$^4$}

\affiliation{$^1$  Laboratoire de Physique Th\'eorique et Hautes Energies, CNRS,
UPMC Univ Paris VI Boite 126, 4
Place Jussieu, 75252 Paris cedex 05, France (UMR du CNRS 7589) }
\affiliation{$^2$Sorbonne Universit\'es, Institut Lagrange de Paris (ILP), 98 bis bd Arago 75014 Paris, France}
\affiliation{$^3$ Raymond and Beverly Sackler
School of Physics and Astronomy \\
Tel-Aviv University, Ramat-Aviv 69978, Israel}
\affiliation{$^4$ Laboratoire de Physique Th\'eorique, Ecole Normale Sup\'erieure, 24 rue
Lhomond, 75231 Paris Cedex 05, France (UMR du CNRS 8549)}

\date{\today}
\begin{abstract}
Supersymmetry is spontaneously broken when the field theory stress-energy tensor
has a non-zero vacuum expectation value.
In local supersymmetric field theories the massless gravitino and goldstino
combine via the super-Higgs mechanism to a massive
gravitino. We study this mechanism in four-dimensional fluids, where the vacuum
expectation value of the stress-energy tensor
breaks spontaneously both supersymmetry and Lorentz symmetry. We consider both constant as
well as space-time dependent ideal fluids.
We derive a formula for the gravitino mass in terms of the fluid velocity,
energy density and pressure.
We discuss some of the phenomenological implications.

\end{abstract}


\maketitle

\tableofcontents

\section{Introduction}

A spontaneous breaking of supersymmetry is manifested by the generation of a massless fermionic Goldstone mode, the \emph{goldstino} \cite{Fayet:1974jb}. At non-zero temperature the vacuum expectation value of the stress-energy tensor
breaks spontaneously supersymmetry as well as Lorentz symmetry, and
the goldstino mode is called \emph{phonino} (see e.g. \cite{Boyanovsky:1983tu,Leigh:1995jw,Kratzert:2003cr}).
We may view the phonino as a (supersymmetric) sound mode. However, unlike the ordinary bosonic sound mode that can be treated
as a classical field, the phonino is fermionic and is therefore inherently a quantum field \cite{Hoyos:2012dh}.
Its dispersion relation at leading order 
in momenta is fixed by the supersymmetry algebra.  It is linear with velocity $v=\left|\frac{p}{\epsilon}\right|$, where $p$ and $\epsilon$ are
the fluid pressure and energy density, respectively.  
When the spontaneous supersymmetry breaking is due to a cosmological constant with the equation of state $\epsilon = -p$, the phonino is
the ordinary goldstino,  whose 
velocity is the speed of light.

 In local supersymmetric field theories the goldstino combines with the massless gravitino via the super-Higgs mechanism to form a massive gravitino 
 \cite{Volkov:1973jd},. 
The aim of this paper is to study the super-Higgs mechanism in four-dimensional ideal fluids.
Supersymmetry breaking is parametrized by the vacuum expectation value (vev) of the ideal fluid stress-energy tensor.
We construct  the effective field theory of the fermionic low-energy modes, the phonino and the gravitino, 
in the background of the fluid stress-energy tensor. We work up 
to quadratic order in the fields and to first order in derivatives. In the following we outline the results.
Consider first the case, where the energy density and the pressure are constant, and the fluid
 is in the rest frame.
Diagonalizing the field equations, we show that the longitudinal mode of the gravitino mixes with the phonino and
acquires a mass
\begin{equation}
m_{gravitino} = \frac{\sqrt{3}}{4 M_p}
\left|\frac{p-\frac{\varepsilon}{3}}{\sqrt{\varepsilon}} \right| \ ,
\label{massgr}
\end{equation}
where $M_p$ is the Planck mass.
The dispersion relation of this mode is inherited from that of the phonino and is non-relativistic. The transverse part of the gravitino
acquires the same mass (\ref{massgr}), however, its dispersion relation is relativistic.

When the spontaneous supersymmetry breaking is due to a cosmological constant $T^{\mu\nu} = -F^2 \eta^{\mu\nu}$, one gets
from (\ref{massgr}) the well known formula for the gravitino mass $\frac{F}{\sqrt{3}M_p}$.
Note, also that the gravitino mass vanishes for a conformal fluid, where the stress-energy tensor is traceless and the equation of state 
is $\epsilon-3p=0$.

We study next the super-Higgs mechanism in the background of a non-constant slowly varying stress-energy tensor.
We derive a general constraint on the gravitino field and analyze in detail the case of time-dependent energy density and
pressure. The mass terms in this case include contributions from derivatives of the energy density.

The paper is organised as follows. In section \ref{basics} we briefly review various aspects of the goldstino, gravitino and
the standard super-Higgs mechanism.
 In section 
\ref{superhiggs} we consider the spontaneous breaking of supersymmetry
and Lorentz invariance  due to a non-zero vacuum expectation value of the fluid
stress-energy tensor. We introduce the phonino field and its couplings to the gravitino, and 
study the super-Higgs mechanism.
We consider first the case of a constant stress-energy tensor, and then extend the analysis
to the case to space-time dependence, working to first order in derivatives.  
In section \ref{idealcase} 
we study  in detail the field equations and find the propagating modes. 
For a constant stress-energy tensor we show that the goldstino is eaten by the gravitino, but retains its identity 
 as it survives as the longitudinal mode of the gravitino with its own
dispersion relation. The transverse and longitudinal component become massive,
with the same mass (\ref{massgr}).
We generalize the discussion and study the field equations and the mass terms 
in the background of time-dependent stress-energy tensor. 
The last section is devoted to a discussion of some phenomenological implications. 

\section{F-term susy breaking and the super-Higgs mechanism}\label{basics}

\subsection{Goldstino and Gravitino}

In a global supersymmetric theory in flat space time, supersymmetry is broken
spontaneously when the vacuum has non-zero 
energy. Preserving Lorentz invariance, this is typically accomplished for $N=1$
susy in 4 dimensions by giving a vev to an auxiliary field in a chiral multiplet
(F-term) or in a vector multiplet (D-terms). 
As a consequence of Goldstone theorem,  the low energy spectrum contains a
fermionic massless mode, known as the $\mathit{goldstino}$.

The goldstino is a spin $\frac{1}{2}$ field $(G_{\alpha},\bar{G}_{\a})$ in the
$(\frac{1}{2},0)\oplus (0,\frac{1}{2})$ representation of the Lorentz group\footnote{We will use Wess
and Bagger notations. $\eta_{\mu\nu} = diag(-,+,+,+)$,
$\epsilon^{12}=-\epsilon^{21}=1$. $\zeta_{\alpha}$ is a left Weyl
spinor in the $(\frac{1}{2},0)$ representation. $\bar{\zeta}_{\a}$ is a right
Weyl
spinor in the $(0,\frac{1}{2})$ representation. Complex conjugation exchanges
$SU(2)_L$ and $SU(2)_R$. The complex conjugate of a left Weyl spinor is a right
Weyl spinor.}.
Its mass dimension is $\frac{3}{2}$. At quadratic order, the Lagrangian that
describes its dynamics is only a kinetic term 
\beq
{\cal L}_{G} =- i \bar{G}\bar{\sigma}^{\mu}\partial_{\mu}G ,
\label{LG}
\eeq
and the field satisfies the Dirac equation
\beq
\bar{\sigma}^{\mu}\partial_{\mu}G = 0,~~~~ \sigma^{\mu}\partial_{\mu}\bar{G} = 0
\ .
\label{feqgold}
\eeq
Theories with $N=1$ local supersymmetry contain a gravitino  field
$(\psi_{\mu\alpha},\bar{\psi}_{\mu\a})$ of spin $\frac{3}{2}$ and mass dimension
 $\frac{3}{2}$. Following Fierz and Pauli, the irreducible spin  $\frac{3}{2}$ 
representation is obtained from 
$\psi_{\mu\alpha}$  in the $(\frac{1}{2},\frac{1}{2})\otimes(\frac{1}{2},0) =
(1,\frac{1}{2})\oplus(0,\frac{1}{2})$ representation, and  $\bar{\psi}_{\mu\a}$ 
in the $(\frac{1}{2},\frac{1}{2})\otimes(0,\frac{1}{2}) =
(\frac{1}{2},1)\oplus(\frac{1}{2},0)$ representation by imposing constraints
that project out  the additional spin $\frac{1}{2}$  components. The
$(0,\frac{1}{2})$ and $(\frac{1}{2},0)$ parts in the decomposition of
$(\psi_{\mu\alpha},\bar{\psi}_{\mu\a})$
are removed by imposing
\beq
\bar{\sigma}^{\mu}\psi_{\mu} = 0,~~~~ \sigma^{\mu}\bar{\psi}_{\mu} = 0 \ .
\label{constraint}
\eeq
The representations $(1,\frac{1}{2})$ and  $(\frac{1}{2},1)$ have dimension six
each. In order to reduce the number of degrees of freedom
to four we impose
\beq
\partial^{\mu}\psi_{\mu\alpha}=0,~~~~ \partial^{\mu}\bar{\psi}_{\mu\a}=0 \ .
\label{constraintder}
\eeq

One can get this structure of equations and constraints from a Lagrangian.
The massless gravitino Rarita-Schwinger Lagrangian is:
\beq
{\cal L}_{\psi} = 
\ep^{\mu\nu\rho\sigma}\bar{\psi}_{\mu}\bar{\sigma}_{\nu}\p_{\rho}\psi_{\sigma} \
.
\eeq
The field equations are
\beq
\ep^{\mu\nu\rho\sigma}\bar{\sigma}_{\nu}\p_{\rho}\psi_{\sigma} = 0,~~~~
\ep^{\mu\nu\rho\sigma}\sigma_{\nu}\p_{\rho}\bar{\psi}_{\sigma} = 0 \ .
\eeq

By imposing on this equation the condition (\ref{constraint}) we get
\beq
\bar{\sigma}^{\rho}\p_{\rho}\psi_{\sigma} = 0,~~~~
\sigma^{\rho}\p_{\rho}\bar{\psi}_{\sigma} = 0 \ .
\label{gravitinoeq}
\eeq
It is easy to see that (\ref{gravitinoeq}) and (\ref{constraint}) imply
(\ref{constraintder}).

\subsection{The superHiggs mechanism}

Consider a spontaneous F-term supersymmetry breaking in a theory with local supersymmetry.
The stress-energy tensor in this case has a vev 
$T^{\mu\nu}  = - F^2 \eta^{\mu\nu}$, where
$F$ is the vev of the auxiliary field. We take $F$ real with mass dimension two.
The supersymmetry transformations (we suppress here the spinor index) are:
\begin{equation}\begin{split}\label{susytran}
\delta \psi_{\mu} = -M_p\left(2\partial_{\mu}\epsilon + n
\sigma_{\mu}\bar{\epsilon}\right),~~~~\delta G = \sqrt{2}F \ep \  , \\
\delta \bar{\psi}_{\mu} = -M_p\left(2\partial_{\mu}\bep +
\bar{n}\bar{\sigma}_{\mu}\epsilon \right),~~~~
\delta \bar{G} = \sqrt{2}F\bep \ .
\end{split}\end{equation}
In order to have a Lagrangian for the gravitino and goldstino invariant under
(\ref{susytran}) we need to add mass terms:
\bea\label{lmass}
{\cal L}_{mass} &=& M^{\mu\nu}\psi_{\mu}\psi_{\nu} +
\bar{M}^{\mu\nu}\bar{\psi}_{\mu}\bar{\psi}_{\nu} +
\left(C^{\mu\nu}\psi_{\mu}\sigma_{\nu}\bar{G} -
\bar{C}^{\mu\nu}\bar{\psi}_{\mu}\bar{\sigma}_{\nu}G \right)
+ \frac{1}{2}\left(mGG  +  \bar{m}\bar{G}\bar{G} \right)\, .
\label{mass}
\eea
$\bar{M}^{\mu\nu}$, $\bar{C}^{\mu\nu}$ and $\bar{m}$ are the complex conjugates
of
$M^{\mu\nu}$, $C^{\mu\nu}$ and $m$, respectively.
We require that the total action
\beq
{\cal L} = {\cal L}_{\psi} + {\cal L}_{G} + {\cal L}_{mass} \ ,
\eeq
be invariant under the supersymmetry transformations (\ref{susytran}).

The supersymmetry invariance of the action implies fixes the action uniquely.
From terms of the form $\p\epsilon\psi$ we get
\beq
M^{\mu\nu} = i\bar{n} \sigma^{\mu\nu} \ .
\eeq
We set $-i\bar{n} = m_{\frac{3}{2}}$, thus, the gravitino mass matrix reads
\beq
M^{\mu\nu}\psi_{\mu}\psi_{\nu} = -
m_{\frac{3}{2}}\psi_{\mu}\sigma^{\mu\nu}\psi_{\nu} \ .
\eeq
From the $\bar{\epsilon}\psi$ terms we get
\beq
C^{\mu\nu} = \frac{i\sqrt{3}}{\sqrt{2}}m_{\frac{3}{2}} \eta^{\mu\nu} \ .
\eeq
The mass of the gravitino is determined from the $\p\epsilon\bar{G}$ terms
\beq
m_{\frac{3}{2}} = \frac{F}{\sqrt{3}M_p} \ .
\label{gravitinomass}
\eeq
From the $\epsilon G$ terms we get 
\beq
m = - m_{\frac{3}{2}} \ .
\eeq

We can read the propagating degrees of freedom most easily by going to the unitary gauge, where we use the susy transformations to set  $G=\bar{G}=0$; then we find the Lagrangian for a massive gravitino 
\beq
{\cal L}_{g} = 
\ep^{\mu\nu\rho\sigma}\bar{\psi}_{\mu}\bar{\sigma}_{\nu}\p_{\rho}\psi_{\sigma} 
 - m_{\frac{3}{2}}\psi_{\mu}\sigma^{\mu\nu}\psi_{\nu} -
m^*_{\frac{3}{2}}\bar{\psi}_{\mu} \bar{\psi}{\sigma}^{\mu\nu}\bar{\psi}_{\nu}  \
 .
\eeq
From the lagrangian we can also read the form of the supercurrents 
$(S_{\mu\alpha},\bar{S}_{\mu\a})$, $\bar{S}_{\mu}^{\a}=
\left(S_{\mu}^{\alpha}\right)^{\dagger}$. 
They couple to the gravitino as
\begin{equation}\label{Spsi}
\frac{1}{M_p}\int d^4 x \left( S_{\mu\alpha}\psi^{\mu\alpha} +  
\bar{S}_{\mu\a}\bar{\psi}^{\mu\a} \right) \ .
\end{equation}
Comparing to (\ref{lmass}) we see that 
\beq
S^{\mu\alpha} = i\frac{F}{\sqrt{2}}\sigma^{\mu} \bar{G},~~~~\bar{S}^{\mu\a} =
-i\frac{F}{\sqrt{2}}\bar{\sigma}^{\mu} G \ .
\label{supercurrents}
\eeq
These expressions are the leading non-derivative terms in the derivative expansion of
the supercurrents.
The conservation laws
\beq
\p_{\mu}S^{\mu\alpha} = 0,~~~~\p_{\mu}\bar{S}^{\mu\a} = 0 \,
\eeq
are the field equations of the goldstino (\ref{feqgold}).

\section{ The Super-Higgs Mechanism in Fluids}\label{superhiggs}

In this section we will study the super-Higgs mechanism in fluids, where the
vacuum expectation value of the stress-energy tensor
$T_{\mu\nu}$ breaks spontaneously both supersymmetry and Lorentz symmetry.
One of the motivations for our study is to understand the fate of the
{\it phonino} in supergravity theories.

\subsection{Supersymmetric fluids}\label{fluid}

 Consider a supersymmetric field theory in thermal equilibrium described by a background stress-energy tensor
\begin{equation}
T^{\mu\nu}={\rm diag}\,(\varepsilon,p,p,p) \ .
\label{T}
\end{equation}
$p$ is the pressure and $\varepsilon$ is the energy density, and the two are
related by an equation of state $p(\varepsilon)$.
The expectation value of the stress-energy tensor (\ref{T}) breaks spontaneously supersymmetry and Lorentz symmetry 
but keeps rotational invariance.
Two special cases of (\ref{T}) are: $-p = \varepsilon = F^2$ corresponding to the F-term breaking, and $p =
\varepsilon/3$ that describes a conformal fluid. 

The spontaneous breaking of supersymmetry implies in general a massless fermionic field in the spectrum called phonino. 
The existence of this mode can be understood as a consequence of a
supersymmetric Ward-Takahashi identity for the supercurrent two-point function:
\begin{equation}\label{WTsusy}
\partial_{\mu}\langle T \{S^{\mu}(x) \bar{S}^{\nu}(y)\}\rangle \sim
\delta^{(4)}(x-y)\langle T^{\nu\rho}\rangle \sigma_{\rho}  \ .
\end{equation}
Going to momentum space and assuming a constant energy-momentum tensor the
correlator has to have a singularity when $k \to 0$. 
With Lorentz invariance one concludes that there
must be a massless fermionic mode. Without Lorentz invariance it is possible to have a singularity without having a massless
particle. This happens for instance in a free theory. In a generic
interacting system it is expected that the massless mode is present
(see e.g. \cite{Kratzert:2002gh} for a discussion of these issues), and we will consider these cases.

The field equations of the phonino take the form
\beq
T^{\mu\nu}\bar{\sigma}_{\mu}\partial_{\nu}G = 0,~~~~
T^{\mu\nu}\sigma_{\mu}\partial_{\nu}\bar{G} = 0 \ .
\label{Tfeqgold}
\eeq
These equations arise from the Lagrangian
\beq
{\cal L}_{G} = -\frac{i}{{\cal T}^{4}}T^{\mu\nu}
\bar{G}\bar{\sigma}_{\mu}\partial_{\nu}G  \ ,
\label{eqGT}
\eeq
where ${\cal T} = | det \, \langle T^{\mu\nu} \rangle |^{\frac{1}{16}}$.
When $T^{\mu\nu} = -F^2 \eta^{\mu\nu}$ the Lagrangian (\ref{eqGT}) reduces to
(\ref{LG})
and the propagator of the phonino becomes that of the usual goldstino.

\subsection{Generalized super-Higgs mechanism}

In the following we will be working with an expansion in powers of  the dimensionless parameter $\frac {\cal T}{M_p}$. 
The effective Lagrangian for the gravitino and phonino at leading order in this expansion reads
\begin{equation}\begin{split} \label{intermL}
{\cal {L}} =& \epsilon^{\mu\nu\rho\sigma} \bar \psi_\mu \bar \sigma_\nu \p_\rho
\psi_\sigma +i D^{\mu\nu} \bar G \bar \sigma_\mu \p_\nu G + i C^{\mu\nu} ( \bar
\psi_\mu \bar \sigma_\nu G + \psi_\mu \sigma_\nu \bar G) \\
 +& \frac{1}{2} G m G + \frac{1}{2} \bar G m^* \bar G 
 +   M^{\mu\nu}_{\rho\tau} \psi_\mu \sigma^{\rho \tau}  
\psi_\nu +  M^{\mu\nu *}_{\rho\tau} \bar\psi_\mu \bar\sigma^{\rho
\tau}  \bar\psi_\nu  \,.
\end{split}\end{equation}
The mass matrices  $m$ and $M^{\mu\nu}_{\rho\tau}$ have supressed spinor indices. We
could have added also a term $M^{\mu\nu} \psi_\mu \psi_\nu$,  however it turns out that it is not allowed by supersymmetry and we omit it.
Note also,  that at leading order  in $\frac{\cal T}{M_p}$ the gravitino has the standard kinetic term. 

The supersymmetry transformations need to be modified to allow for Lorentz
violating coefficients:
\bea \label{susytransf}
\delta G^{\boldsymbol{\alpha}} &=& \sqrt{2} {\cal T}^2 \ee^{\boldsymbol{\alpha}} \ ,
\nonumber \\
\delta \psi_{\mu \boldsymbol{\alpha}} &=& - M_P ( 2 \p_\mu
\ee_{\boldsymbol{\alpha}} +i  n_{\mu\nu}
\sigma^\nu_{\boldsymbol{\alpha \dot\alpha}} \bar \ee^{\boldsymbol{\dot\alpha}} ) \ , 
 \\
\delta \bar\psi_{\mu \boldsymbol{\dot\alpha}} &=& - M_P ( 2 \p_\mu
\bar\ee_{\boldsymbol{\dot\alpha}} -i 
n^*_{\mu\nu} \ee^{\boldsymbol{\alpha}} \sigma^\nu_{\boldsymbol{\alpha
\dot\alpha}}  )  \ .\nonumber \
\eea
The requirement that the Goldstino equation of motion reproduces, at the lowest
order, the phonino dispersion relation
fixes:
\beq
D^{\mu\nu} =  \frac{T^{\mu\nu}}{ {\cal T}^4}  \ .
\eeq
Note, that this is the 
Volkov-Akulov standard leading term describing the coupling between
matter and Goldstinos, where the stress-energy  tensor appears explicitly with
its non-vanishing vacuum expectation value. We also assumed 
that the supersymmetric vacuum is obtained in flat space when the stress-energy tensor vanishes.

Performing the supersymmetry variation,  terms of the form $\bar G \p_\nu
\ee_{\boldsymbol{\alpha}}$  coming from $\bar G \del G$ and $\bar \psi G$ fix:
\beq
C^{\mu\nu} = - \frac{1}{\sqrt{2}}  \frac{{\cal T}^2}{M_P }  \frac{T^{\mu\nu}}{
{\cal T}^4} \ .
\eeq
This is consistent with a gravitino-phonino coupling of the form (\ref{Spsi}) if
the supercurrent has the form 
\beq
S^{\mu \boldsymbol{\alpha}} \sim  \frac{T^{\mu\nu}}{ {\cal T}^2}
\sigma^{\boldsymbol{\alpha \dot\alpha}}_\nu \bar{G}_{\boldsymbol{ \dot\alpha}} \ .
\eeq
As in the previous section, the conservation equation for the supercurrent is
equivalent to the propagation equation for the phonino. 

Terms of the form $\psi \del \epsilon$ coming from $\bar \psi \del \psi$ and
$\psi \psi$ give 
\beq 
M^{\mu\nu}_{\lambda\kappa}  \sigma^\lambda  \bar \sigma^\kappa= - \frac{i}{2} 
\epsilon^{\mu\nu\rho\sigma} \sigma_\rho   \bar \sigma^\gamma
 n^*_{\sigma \gamma}  \ ,
\eeq 
and terms $\psi \epsilon$ from $\bar\psi G$ and $\psi\psi$ 
\beq 
 M^{ \mu\nu}_{\rho\tau}  n_{\nu\lambda}    \sigma^{\rho \tau}  \sigma^\lambda
  =  -\frac{T^{\mu\nu}} {2 M_P ^2}   \sigma_{\nu} \,.
\eeq

The last two equations lead to:
\beq 
 \frac{i}{2}  \epsilon^{\mu\nu\rho\sigma} n^*_{\nu\lambda} n_{\sigma \gamma}
\sigma_\rho\bar\sigma^\gamma \sigma^\lambda 
 =  \frac{T^{\mu\nu}} {M_P ^2}   \sigma_{\nu} \ .
\eeq
The last equation can be put in a simpler form when $n$ is real, which we will assume from now on. 
We 
antisymmetrize in $\rho\gamma\lambda$ and get
\beq
-\half \epsilon^{\mu\nu\sigma\rho} \epsilon_{\rho}^{~\lambda\gamma\kappa}
n_{\nu\lambda} n_{\sigma\gamma} = 
\frac{T^{\mu\kappa}}{M_P^2} \ .
\label{nT}
\eeq
This equation determines $n_{\mu\nu}$ in terms of $T_{\mu\nu}$. 
Finally, from the terms $\ee G$ we have:
\beq 
{ m_{\boldsymbol{\alpha}}}^{\boldsymbol{\beta}}=  \frac{1}{2} \frac{T^{\mu\nu}
n_{\mu\nu}} {{\cal T}^4}  {\delta_{\boldsymbol{\alpha}}}^{\boldsymbol{\beta}} \ .
\eeq
To arrive at this form we used the fact that $T^{\mu\rho} n_{\rho}^{\nu}$
is symmetric in $\mu\nu$. 
Putting all the results together gives the Lagrangian
\bea\label{finalL}
{\cal {L}} &=& \epsilon^{\mu\nu\rho\sigma} \bar \psi_\mu \bar \sigma_\nu \p_\rho
\psi_\sigma  + \frac{i}{4}  \epsilon^{\mu\nu\rho\sigma}  n_{\sigma \gamma} 
\bar\psi_\mu  \bar \sigma_\rho   \sigma^\gamma
\bar\psi_\nu - \frac{i}{4}  \epsilon^{\mu\nu\rho\sigma}  n_{\sigma \gamma} 
\psi_\mu  \sigma_\rho   \bar \sigma^\gamma
\psi_\nu \nonumber \\
 &-& \frac{i}{\sqrt{2}}  \frac{{\cal T}^2}{M_P }  \frac{T^{\mu\nu}}{ {\cal T}^4}
( \bar
\psi_\mu \bar \sigma_\nu G + \psi_\mu \sigma_\nu \bar G) \nonumber\\
 &+& i \frac{T^{\mu\nu}}{ {\cal T}^4}  \bar G \bar \sigma_\mu \p_\nu G + 
\frac{1}{4} \frac{T^{\mu\nu} n_{\mu\nu}} {{\cal T}^4}G  G +  \frac{1}{4}
\frac{T^{\mu\nu} n_{\mu\nu}} {{\cal T}^4}\bar G  \bar G  \,.\nonumber 
\eea 

The unitary gauge is obtained by making a supersymmetry transformation to set $G=0$: 
\bea\label{Unitary}
\psi_{\mu \boldsymbol{\alpha}} \rightarrow \psi_{\mu \boldsymbol{\alpha}}  + 
\frac{\sqrt{2}M_P}{{\cal T}^2} \p_\mu G_{\boldsymbol{\alpha}} + i  
\frac{M_P}{\sqrt{2}{\cal T}^2}n_{\mu\nu}
\sigma^\nu_{\boldsymbol{\alpha \dot\alpha}} \bar G^{\boldsymbol{\dot\alpha}}  \,.
\eea 
The resulting Lagrangian reads
\beq
{\cal {L}} = \epsilon^{\mu\nu\rho\sigma} \bar \psi_\mu \bar \sigma_\nu \p_\rho
\psi_\sigma  -\frac{i}{2}  \epsilon^{\mu\nu\rho\sigma}  n_{\sigma}^{~\gamma}
\bar \psi_\mu \bar \sigma_{\rho\gamma}
\bar\psi_\nu +\frac{i}{2}  \epsilon^{\mu\nu\rho\sigma}  n_{\sigma}^{~\gamma} 
\psi_\mu  \sigma_{\rho \gamma} \psi_\nu \,.
\label{gravitinoL}
 \eeq 
The equation of motion is
\beq\label{EOM}
\epsilon^{\mu\nu\rho\sigma} \bar \sigma_\nu \p_\rho
\psi_\sigma  -\frac{i}{2}  \epsilon^{\mu\nu\rho\sigma}  n_{\sigma \gamma}  
\bar  \sigma_\rho   \sigma^\gamma
\bar\psi_\nu = 0 \,.
 \eeq 

Consider now the constraints that are necessary in order to reduce the number of 
degrees of freedom of $\psi_\mu$ to the four that describe a massive gravitino.
 Acting on the equation of motion by $n_{\mu \lambda}  \sigma^\lambda $ gives
\beq
  -\frac{i}{2}  \epsilon^{\mu\nu\rho\sigma} n_{\mu \lambda}  n_{\sigma \gamma}
\sigma^\lambda    \bar  \sigma_\rho   \sigma^\gamma
\bar\psi_\nu = 0 \,.
 \eeq 
 Using the symmetry of $n_{\mu \lambda} $, this can be put in the form:
\beq\label{cs2}
 T^{\mu\nu}  \sigma_\mu \bar\psi_\nu = 0 \ ,
 \eeq 
which replaces the standard F-term breaking constraint $\bar\sigma^\mu \psi^\mu=0$ of the
gravitino.  As in the case of curved space-time \cite{Corley:1998qg}, a second
constraint is obtained  by taking the component $\mu=0$ of (\ref{EOM}). 
We analyze the consequences of the constraint in the next section. 

Consider next the general case of a space-time dependent stress-energy tensor. In the hydrodynamic
regime the fluid is in local thermal equilibrium. One can use a 
derivative expansion since the charge densities 
are slowly varying functions of the space-time coordinates.
At leading order the gravitino Lagrangian in the unitary gauge takes the form
(\ref{gravitinoL}) with 
$n_{\mu}^{\nu}(x^{\alpha})$. As a consistency check we take
the susy variation of the Lagrangian (\ref{gravitinoL}). It yields  
\beq\label{firstordvar}
\delta {\cal L} = - i M_P \,  \epsilon^{\rho\mu\nu\sigma} \, \del_\rho
n_{\mu}^{~\tau} \, \varepsilon \sigma_{\tau} 
\bar\sigma_{\nu}  \psi_\sigma + \frac{i}{\sqrt{2}{\cal T}^2} \del_\nu ({\cal
T}^4 D^{\mu\nu}) \, \bar G \bar \sigma_\mu \ee  \ ,
\eeq
the second term vanishes by the stress-energy tensor conservation. 
The variation (\ref{firstordvar}) can be compensated by adding new terms : 
\bea\label{firstordterms}
{\cal L}^{(1)} &=& i \frac{M_P}{\sqrt{2} {\cal T}^2} \, 
\epsilon^{\rho\mu\nu\sigma} \, \del_\rho n_{\mu}^{\tau} \, \left[ G \sigma_\tau
\bar\sigma_\nu \psi_\sigma + h.c. \right.\nonumber \\ 
 &+& \left. \frac{M_P}{\sqrt{2} {\cal T}^2} G \sigma_\tau \bar\sigma_\nu
\del_\sigma G +  \frac{M_P}{\sqrt{2} {\cal T}^2}\bar G \sigma_\tau
\bar\sigma_\nu \del_\sigma \bar G \right. \\ 
 &+& \left.
i \frac{M_P}{\sqrt{2} {\cal T}^2} n_{\sigma}^\lambda \left(G \sigma_\tau \bar
\sigma_\nu \sigma_\lambda \bar G + i G  \epsilon_{\nu\lambda\tau \gamma}
\sigma^\gamma \bar G\right) \right] \,. \nonumber
\eea
Using the equation (\ref{nT}), the last term can put in the form 
$- \frac{i}{2 {\cal T}^4} \del_\nu T^{\mu\nu} \, \bar G \bar \sigma_\mu G$, so it vanishes 
when  $T^{\mu\nu} $ is conserved\footnote{The stress-energy tensor is conserved when studying a closed system, but we could also consider non-conserved stress-energy tensors, for instance if we apply our formalism  to systems subject to an external force.}. All terms in (\ref{firstordterms}) contain $G$, so they vanish in the unitary gauge,  giving back, at this leading order in the varying mass term, the lagrangian
(\ref{gravitinoL}) and field equations (\ref{EOM}). However, there is a new constraint that replaces
(\ref{cs2}) and takes the form
\beq\label{constr2}
\frac{T^{\mu\nu}}{M_P^2} \sigma_\mu \bar \psi_\nu - \epsilon^{\mu\nu\rho\sigma}
\, \del_\mu n_{\sigma}^{~\gamma} \sigma_\rho \bar\sigma_\gamma \psi_\nu = 0  \,.
\eeq

\section{Ideal fluid}\label{idealcase}

 We will consider now  relativistic ideal fluids with stress-energy tensor 
 \begin{equation}
T^{\mu\nu} = (\epsilon+p)u^{\mu}u^{\nu} + p \eta^{\mu\nu} \ ,
\label{ideal}
\end{equation}
where $u^{\mu}$ is the fluid four-velocity $u^{\mu}u_{\mu} = -1$.
 We will derive the gravitino mass as a function of 
 the fluid variables.
 
In order to solve (\ref{nT}) we parametrize the solution $n_{\mu\nu}$ as
\beq
n_{\mu\nu} = (n_T- n_L) u_\mu u_\nu + n_T \eta_{\mu\nu} \ .
\eeq 
Plugging $n_{\mu\nu}$ and $T_{\mu\nu}$ and solving for $n_T$ and $n_L$ we get 
\begin{equation}\label{solconstr1}
n_T^2 = \frac{\varepsilon} {3 M_P ^2}  \, , \qquad 
-n_T (n_T + 2 n_L) = \frac{p}{M_P^2}  \,,
\end{equation}
hence
\beq \label{solconstr2}
n_L = - n_T \, \left( \frac{\varepsilon + 3 p}{2 \varepsilon} \right) \ .
\eeq 
For $F$-term breaking, $\varepsilon = -p, n_L =  n_T$, and for a CFT, 
$\varepsilon = 3 p, n_L = -  n_T$.

\subsection{Constant stress-energy tensor}

In the following we study in detail the gravitino equations and constraints when
the energy density and pressure are constant and
the 4-velocity is $u^{\mu} = (1,0,0,0)$. In this case $n_{\mu}^{\nu} =
diag(n_L,n_T,n_T,n_T)$. We introduce the notation
\begin{equation}
\slash\!\!\!\!D = \sigma^\mu
\del_\mu,  ~~~~~\dslash= \ \sigma^i \del_i \ ,
\end{equation}
and
\begin{equation}
\Psi= \bar \sigma^\mu\psi_\mu,~~~~{\psi}_{\frac{1}{2}} = \bar\sigma^i \psi_i,~~~~ 
\bar{\psi}_{\frac{1}{2}} = \sigma^i \bar\psi_i  \ .
\end{equation}
 One has
\begin{subequations}
\begin{align} \label{cos0a}
\epsilon^{ijk}\bar \sigma_i \del_j \psi_k = i \bar \sigma^0 ( \dslash  
\bar{\psi_{\frac{1}{2}}} + \del \cdot \psi) \\ 
\label{cos0b}
\epsilon^{0\nu\rho\sigma}  n_{\sigma}^{~\gamma} \bar \sigma_{\rho\gamma}
\bar\psi_\nu = i n_T \bar\sigma^0 \bar{\psi}_{\frac{1}{2}} \ .
\end{align}
\end{subequations}

We rewrite the gravitino equation in the following
form
\beq\label{RS}
\slash\!\!\!\! \bar D  \psi_\mu - \bar \sigma_\mu \del_\nu  \psi^\nu - \del_\mu
\Psi - \bar\sigma_\mu \, \slash\!\!\!\!D \Psi + \epsilon_\mu^{~\nu\rho\sigma}
n_\sigma^{~\gamma} \bar\sigma_{\rho\gamma} \bar \psi_\nu = 0 \ .
\eeq	  
The constraint  (\ref{cs2}) can be used to solve for one of the components 
\bea\label{cons2}
\psi_0 &=& - v \, \sigma_0  \bar{\psi}_{\frac{1}{2}}  \ ,
\eea 
where $v=\left|\frac{p}{\epsilon}\right|$ is the phonino velocity.
The component $\mu=0$ of equation (\ref{RS}) gives the constraint
\bea\label{cons1}
\dslash {\psi}_{\frac{1}{2}}} - i n_T {\bar{\psi}_{\frac{1}{2}}+ \del\cdot\psi
&=& 0 \ .
\eea
Putting all the constraints together leads to  
\bea\label{eqlong}
(\bar \sigma^0 \del_0 + v \, \bar\dslash ) \bar{\psi}_{\frac{1}{2}}  - i \hat m 
{\psi}_{\frac{1}{2}} &=&0 \,.
\eea 
This is the Dirac equation satisfied by the longitudinal spin-1/2 mode  with
mass 
\beq\label{masslong}
\hat m = \frac{n_L+n_T}{2} = \frac{n_T}{4}\left| (1-3v)\right|= \frac{\sqrt{3}}{4 M_P}
\left|\frac{p-\frac{\varepsilon}{3}}{\sqrt{\varepsilon}}\right| \,. 
\eeq
Notice that the eqs. (\ref{solconstr1}) determine $n_L,n_T$ only up to a sign; we have used this freedom in the last equation to 
have a positive mass $\hat m$. 

Using (\ref{cons2}) and (\ref{cons1}) one finds 
\beq\label{scalars}
\Psi = (1+v){\psi}_{\frac{1}{2}} \, , \quad \del_\mu \psi^\mu = (v^2-1) \dslash
{\psi}_{\frac{1}{2}} + i (n_T + \hat m v ) \bar{\psi}_{\frac{1}{2}} \ .
\eeq

Finally, using the last relations in the equation of motion with $\mu=j$ gives 
\beq\label{EOMi}
(\bar\sigma^0 \del_0 + \bar \dslash) \psi_j +i \hat m \bar \psi_j - (1+ v)
\left( \partial_j \bar{\psi}_{\frac{1}{2}} + i \frac{n_T}{2}  \bar \sigma_j 
{\psi}_{\frac{1}{2}} \right) =0  \ .
\eeq

One can verify that contracting this equation with  $\sigma^j$ gives back the
equation for ${\psi_{\frac{1}{2}}}$. 

The projector on the transverse part of the spinor is 
\bea
\psi_j &=& \psi_j^T - \left( \frac{1}{2} \sigma_j - \frac{k_j \sla{k}}{2 k^2}
\right) \bar{\psi}_{\frac{1}{2}} + \left( \frac{3 k_j}{2 k^2}+ \half
\frac{\sigma_j \sla{\bar k}}{k^2}  \right) \, k\cdot \psi \ .
\eea 
Replacing $k \cdot \psi$ using (\ref{cons1}) we have 
\begin{equation}\begin{split}
\psi_j =& \psi_j^T - \frac{k_j \sla{k}}{k^2} \bar{\psi}_{\frac{1}{2}} +
\frac{n_T}{2 k^2} \left( k_j - \sla{k} \bar\sigma_j \right)
{\psi_{\frac{1}{2}}}  \\
\bar \psi_j =& \bar \psi_j^T - \frac{k_j \sla{\bar k}}{k^2} 
{\psi_{\frac{1}{2}}} - \frac{n_T}{2 k^2} \left( k_j - \sla{\bar k} \sigma_j
\right) \bar{\psi}_{\frac{1}{2}} \ .
\end{split}\end{equation}
One can check plugging it in (\ref{EOMi}) that the transverse part satisfies
the decoupled equation 
\beq \label{eqtrans}
(\bar\sigma^0 \del_0 + \bar \dslash) \psi_j^T +i \hat m \, \bar \psi_j^T = 0 \ .
\eeq

Eqs. (\ref{eqlong}),(\ref{masslong}) and (\ref{eqtrans}) are our main results.
In the fluid, the goldstino is eaten by the gravitino. The gravitino 
has two distinct propagating modes, the longitudinal and the transversal,
with the same mass but different dispersion relations. 
It is interesting to note that the gravitino and the goldstino remain massless
in a CFT fluid. 

\subsection{Slowly varying stress-energy tensor} 

Consider and ideal fluid with time dependent stress-energy, $\epsilon=\epsilon(t), p = p(t)$.
 The field equation (\ref{RS}) is  still valid, however 
the constraint (\ref{cons1}) now reads 
\beq 
\frac{\epsilon}{M_P} \sigma_0 \bar\psi_0 + \frac{p}{M_P} \sigma_k \bar\psi_k - i
\frac{\dot\epsilon}{\sqrt{3  \epsilon}} \sigma_0 \bar\sigma_k \psi_k = 0 \ .
\eeq 

It is straightforward to solve the constraints and derive the mass for the
longitudinal and transverse components. 
First notice that (\ref{cos0a},\ref{cos0b},\ref{RS},\ref{cons1}) are unchanged, while for $\psi_0$ we get 
\beq\label{cons2bis}
\bar\psi_0 = - v \, \bar\sigma_0 \bar{\psi}_{\frac{1}{2}}  + i M_P
\frac{\dot\epsilon}{\sqrt{3} \epsilon^{3/2}} {\psi_{\frac{1}{2}}} \ , 
\eeq 
and the equation for the longitudinal mode becomes 
\beq
(\bar \sigma^0 \del_0 + v \, \bar\dslash ) \bar{\psi}_{\frac{1}{2}} -  i M_P
\frac{\dot\epsilon}{\sqrt{3} \epsilon^{3/2}} \bar\sigma^0 \dslash
{\psi_{\frac{1}{2}}}  - i \hat m  \bar{\psi}_{\frac{1}{2}} +
\frac{\dot\epsilon}{2 \epsilon} \bar\sigma^0 \bar{\psi}_{\frac{1}{2}} = 0 \  .
\eeq
In comparison  to the time-independent case, we see that both the dispersion relation and the mass are modified by the time-dependent terms, and couple the different chiralities. These general features are in agreement with the results of \cite{Giudice:1999yt,Kallosh:2000ve} though the details differ. We comment on these differences in the final section. 

Equation (\ref{scalars}) takes the form 
\begin{equation}\begin{split}
\Psi =& (1+v) {\psi_{\frac{1}{2}}} - i M_P \frac{\dot\epsilon}{\sqrt{3}
\epsilon^{3/2}} \bar\sigma^0 \bar{\psi}_{\frac{1}{2}}  \,,\\
\del_\mu \psi^\mu =& (v^2-1) \dslash {\psi_{\frac{1}{2}}} + i (n_T + \hat m
v ) \bar{\psi}_{\frac{1}{2}} - \hat m M_P \frac{\dot\epsilon}{\sqrt{3}
\epsilon^{3/2}} \sigma^0 {\psi_{\frac{1}{2}}} \ .
\end{split}\end{equation}
Considering (\ref{RS}) for $\mu=j$, one sees that equation of motion
for the transverse part will remain unchanged.

\section{Discussion}

Our results can find diverse phenomenological applications. 
One can 
consider Standard Model particles  as part of the fluid under
consideration. This would be the situation during early  universe evolution, and it has 
been studied in \cite{Kallosh:1999jj},
\cite{Giudice:1999yt},\cite{Giudice:1999am},\cite{Kallosh:2000ve} using the
the framework of $N=1$ supergravity  in a FRW background that arises as a solution to Einstein
equations with the fluid stress-energy tensor. Although our framework is different
our results for the gravitino field equations agree with theirs upon making a number of identifications.
For instance, in our formulae the stress-energy momentum tensor contains both
the contribution of the fluid and the hidden sector responsible through its 
$F$-term, $F=\sqrt{3} m_{\frac{3}{2}} M_p$ of the supersymmetry breaking at zero
temperature, $T^{\mu\nu} \rightarrow T^{\mu\nu}-F^2
\eta^{\mu\nu}$ as well as proper rescaling by the vierbein.
The case of
varying stress-energy momentum tensor can not be compared directly as it is 
based on different assumptions.

Another phenomenological application is to identify the fluid as a
hidden sector. The supersymmetry breaking is mediated through very weak
interactions not sufficient to thermalize the whole system. For instance
gravitational interactions will lead to soft terms of the order of 
$m_{soft} \sim \frac {{\cal T}^2}{M_p}$. The mediation also induces a Lorentz
violation in the visible sector, therefore implying a bound on $1-\frac{v}{c}$
for  the viability of this scenario \cite{progress}. 

Beyond the original motivation of studying supersymmetry breaking, the super-Higgs mechanism allows to engineer
Lagrangian for massive  Rarita-Schwinger fields that do not exhibit
pathologies such as breakdown of causality \cite{Velo:1969bt}. In this way, our 
Lagrangian, and the corresponding  equations of motion, can be thought of as describing
the propagation of a spin $\frac{3}{2}$ state, e.g. a  hadronic
resonance,  in a non-Lorentz invariant background.

Finally, it is of interest to continue the study of the hydrodynamics of supersymmetric field theories \cite{Hoyos:2012dh,Kovtun:2003vj}
beyond the ideal order 
(see \cite{Policastro:2008cx, Gauntlett:2011wm,Kontoudi:2012mu,Erdmenger:2013thg} for a computation of transport coefficients at strong coupling using AdS/CFT).  
The study of couplings of the supersymmetry hydrodynamic modes to the gravitino is a useful framework to pursue, at least for the 
analysis of non-dissipative transports.

\section*{Acknowledgements}

We would like to thank Luc Darm\' e for discussions and comments on the manuscript. 
We  would like to thank the CERN theory unit where part of this work was done.
Y.O. would like to thank Laboratoire de Physique Th\'eorique et Hautes Energies 
and Ecole Normale Sup\'erieure, snd G.P. the University of Tel Aviv, for their hospitality.
This work is supported in part by the Israeli Science Foundation Center
of Excellence, and by the I-CORE program of Planning and Budgeting Committee and the Israel Science Foundation (grant number 1937/12)
and in part by French state funds managed by the ANR Within the Investissements d'Avenir programme under reference ANR-11-IDEX-0004-02.

\end{document}